# Online Channel Assignment in Multi-Radio Wireless Mesh Networks Using Learning Automata

Ziaeddin Beheshtifard
Department of Computer engineering and IT
Islamic Azad University, Qazvin Branch
Qazvin, Iran
beheshti@qiau.ac.ir

Mohammad Reza Meybodi
Department of Computer engineering and IT
Amirkabir University of Technology
Tehran, Iran
mmeybodi@aut.ac.ir

*Abstract*— In this paper, we look into the problem of channel assignment in multi-channel multi-radio wireless mesh networks. We propose a new learning automata based channel assignment scheme that adaptively improve network overall throughput by expecting channel state. Since the ability of sending packets via upstream links will be evaluation bases for assigning channels to radio interfaces on each node. We use a link capacity function that potentially reflects degree of interferences imposed by selected channels by each node. According to dynamics of system, proposed algorithm assigns channels to radio interface in distributed fashion such that minimize interference in neighborhood of a node. We analyze the stability of the system via appropriate Lyapunov-like trajectory; we show that stability and optimum point of the system is converged.

*Keywords—Learning Automata; Wireless Mesh Networks; Channel Assignment*

## I. INTRODUCTION

A wireless mesh network (WMN), consists of mesh routers and mesh clients. The mesh routers are generally stationary nodes and form a multi-hop wireless backbone (referred to as the backhaul tier) between the mesh clients and the Internet gateways (a gateway is the node directly connected to the wired network). Each mesh router operates not only as a host but also as a router, forwarding packets on behalf of other nodes that may not be within direct wireless transmission range of their destinations. Additionally some of mesh routers operate as gateways that interconnect wireless backbone to wired internet backbone. On the other hand, mesh clients form the client tier. They are either stationary or mobile, and can form a client mesh network with each other and with mesh routers. The gateway and bridge functionalities in mesh routers enable the integration of WMNs with various existing wireless networks such as wireless sensor, cellular, wireless-fidelity (Wi-Fi), and worldwide inter-operability for microwave access (WiMAX).

WMNs are noticed as a highly reliable, compromising and low cost solution for wireless networks where cover large areas internet access through multi-hop communications. Alongside these advantages, other features like scalability and fast deployment of these networks made it increasingly popular. The networking solution based on WMNs mitigates many of the disadvantages of the conventional WLAN architecture based on a digital subscriber line (DSL) where the hop is wireless. For example, within the WLAN scenario, even if information has to be shared within a community or neighborhood, all traffic must flow through the Internet. Moreover, only a single path may be available for one house to access the Internet. Additionally, wireless services must be set up individually at every home. As a result, network service costs may increase [1]. Deployment of a WMN is a robust and inexpensive alternative; the wireless backbone has the ability to support both internal (among mesh routers) and external (to the Internet) traffic. It also guarantees the existence of multiple paths and makes it possible to cover larger areas with lower costs.

In this paper, we first summarize the related studies on channel assignment problem in section II. Next we focus on system model that consist of stochastic learning method used in learning automata and propose a distributed multi radio channel assignment based on leaning automata that gradually learn stochastic behavior of network in section III and IV. After that we map the network model to a game of multi-automata model and analyze the stability of the proposed scheme in section V. The rest of the paper, provide some simulation result and conclusions.

## II. CHANNEL ASSIGNMENT PROBLEM

A node in wireless mesh networks needs to share a common channel with its neighbors. Assigning more radios to the same channel, results the more connectivity achievement. Obviously establishing more virtual links with neighbors, increases interference. Therefore, there exist a tradeoff between maximizing connectivity and minimizing interference [21]. Increase in neighbors who use the same channel can increase interference. So we need a scheme that creates balance between maximizing the connections and reduction of interference.

Restrictions that a channel allocation algorithm is faced with are as follows: First, total number of channels is fixed. Second, Number of separate channels for each mesh router limit to the number of radios that it has. Third, two nodes that require communicating with each other, need to share a common channel. Forth, total traffic on a connection that passes from a common channel should not go higher than the nominal capacity of channel [2].

In general, all of channel assignment approaches are categorized into two main classes: centralized or distributed. A good classification of these approaches has studied in [20]. According to this survey, most of centralized approach uses graph based techniques to solve problem with mentioned constraints. CLICA is an approach that proposed by Marina et al [11]. That is called also base channel assignment keep any link of unit disk graph and according to distributing of non-overlapping channels among different links, the interference is reduced and finally the traffic pattern was not considered [11]. This algorithm mainly has some limitations. By using the unit disk graph to assign channels, it is difficult to model the number of radios at each node, so in the end of the algorithm, an additional step is required to handle these unassigned radios. Additionally fairness may be sacrificed because it picks a channel for a link in a locally optimal fashion. INSTC [12], proposed by Tang et al., is similar to CLICA therein it also models the CA problem into the problem of assigning channels to links in the unit disk graph. The basic idea of the INSTC algorithm is to assign the channels by traversing the links in a k-connected sub-graph of the given unit disk graph in a predetermined order. Both of the CLICA and INSTC have similar behaviors in their algorithms, where is that they both pick a channel for a link in a locally optimal manner, but are different in that INSTC uses a predetermined order to traverse the links, while CLICA dynamically adjusts its traversal order according to the degree of flexibility obtained from the constantly updated conflict graph. According to similarity of these two approaches, INSTC limitation is similar to CLICA.

In term of learning method, our work is similar to the approach presented in [16] but different in channel state expectation. In [16], the authors proposed an approach that each mesh node discovers its neighbors and the channel usage in its neighborhood periodically. In our scheme, nodes do not need to pass messages with neighbors, and independently expect channel state according to the ability of packet transmission.

In some studies, channel assignment problem modeled by graph coloring problem, but this approach could not satisfied all of constraint. For instance, standard graph coloring does not consider the all of mentioned constraint [3,13]. Node multi-coloring could not resolve common channel constraint [14]. On the other hand edge coloring violate the second constraint, such that number of assigned color to each node cannot exceed the number of radios that each node has. Although constrained edge coloring scheme covers first tree constraint but could not resolve traffic constraint.

This paper addresses the channel assignment problem and specifically investigates optimal channel assignment in wireless mesh networks using learning automata. In this approach, nodes do not require any knowledge about network topology and heuristically learn contention for better channel selection in distributed fashion. Our algorithm intelligently selects channels for the mesh radio in order to minimize interference between mesh routers and maximize connectivity between them. On the other hand this algorithm tries to minimize ripple effect, that decrease throughput of system because of channel oscillation.

### III. SYSTEM MODEL

This section presents an overview of learning automata as an approach for dynamic optimization for stochastic behavior of wireless mesh networks. This unknown behavior affects some network conditions such as distortion and fading are unknown in considered real-time applications. The online algorithms are implemented in real-time in order to cope with unknown application characteristics, network dynamics and resource constraints. This section also presents problem formulation for channel assignment problem.

#### A. Learning Automata

A learning automaton is an adaptive decision-making mechanism situated in a random environment that learns the optimal action through repeated interactions with its environment. Our discussion focuses on a specific type of learning automata called variable structure stochastic learning automata. The stochastic automaton attempts a solution of the problem without any information on the optimal action (initially, equal probabilities are attached to all the actions). One action is selected at random, the response from the environment is observed, action probabilities are updated based on that response, and the procedure is repeated. A stochastic automaton acting as described to improve its performance is called a learning automaton.

The operation of a LA is based on the probability updating algorithm, also known as the reinforcement scheme. This algorithm uses the environmental response that was received as a result of performing the action selected at cycle n (action $a(n)$), in order to update the probability distribution vector p. After the updating is performed, the LA selects the action to perform at cycle $n+1$, according to the updated probability distribution vector $p(n+1)$. The Learning algorithm of the automata can be obtained from ordinary differential equation (ODE) that well approximates the behavior of the algorithm. The stochastic algorithm can be state by following update equation:

$$p_{k+1} = p_k + bG(p_k, \xi_k) \quad (1)$$

Where $p_k \in \Re^N$, is called the state vector and $\xi_k \in \Re^{N'}$, is the noise vector and $b$ is step size or learning parameter.

A general reinforcement scheme has the form of Equations (2,3):

$$p_i(n+1) = p_i(n) - (1-\beta(n))g_i(p(n)) \\ -\beta(n)h_i(p(n)) \quad if \ \alpha(n) \neq \alpha_i \quad (2)$$

$$p_i(n+1) = p_i(n) - (1-\beta(n))\sum_{j\neq i}g_i(p(n))$$
$$-\beta(n)\sum_{j\neq i}h_i(p(n)) \quad if \ \alpha(n)=\alpha_i \quad (3)$$

The functions $g_i$ and $h_i$ are associated with reward and penalty for the selected action $a_i$, respectively, while $\beta(n)$ is a parameter expressing the received environmental response at cycle n, normalized in the interval [0,1]. The lower the value of $\beta(n)$, the more favorable the response is.

According to functions $g$ and $h$, different schemes of reinforcement can be used. The simplest and most commonly used of them is linear schemes. They include the linear reward-penalty ($L_{R-P}$), and linear reward-inaction ($L_{R-I}$). The general linear reward-penalty can be described as (4) – (6). If $\alpha(k)=\alpha_i$ then $p(k)$ is updated as

$$p_i(k+1) = p_i(k) + \lambda_1\beta(k)(1-p_i(k)) - \lambda_2(1-\beta(k))p_i(k)$$
$$p_j(k+1) = p_j(k) - \lambda_1\beta(k)p_j(k)$$
$$+ \lambda_2(1-\beta(k))\left(\frac{1}{r-1} - p_j(k)\right), j\neq i \quad (4)$$

The Linear Reward-Inaction ($L_{R-I}$) algorithm updates the action probabilities as described below:

$$p_i(k+1) = p_i(k) + \lambda\beta(k)(1-p_i(k))$$
$$p_j(k+1) = p_j(k) - \lambda\beta(k)p_i(k) \quad \forall j \neq i \quad (5)$$

The above algorithm can be rewritten in vector notation as

$$P(k+1) = P(k) + \lambda\beta(k)(e_i - P(k)) \quad (6)$$

Here $e_i$ is the unit vector with $i^{th}$ component unity where the index $i$ correspond to the action selected at $k$ and $\lambda$ is the learning (or step-size) parameter satisfying $0<\lambda<1$. Compared to $L_{R-P}$, the main difference here is that, even the reinforcement is unfavorable (that is, $\beta(k)=0$) the action probabilities are not updated.

### B. Problem Formulation

Consider a wireless mesh network with N nodes and L links in mesh routers backbone. Each link corresponds to a pair of transmitter node and receiver. Let $C=\{c_1,c_2,...c_k\}$ is set of available distinct channel in the system. $b(l)$ and $e(l)$ denote the transmitter node and the receiver node, respectively, of link $l$ and $V(v_i)$ is a set of nodes that are in interference range of node $v_i$. We define vector L such that:

$$L(v_i) = \{l_j : v_i = b(l_j) \ or \ v_i = e(l_j)\} \quad (7)$$

Also we use $x_c(l)$ that represent rate at link l can transfer data on channel c, when there are no interfering links transmitting on channel c at the same time. We can define a set $I_l$ of links that interfere with link l as below:

$$I_l = \{l' : b(l') \in V(b(l))\} \quad (8)$$

Link $l$ and another link $l'$ can transmit at the same time with different channels. This model of interference is very general and can be used in a large class of interference relationships, including IEEE 802.11DCF. We consider each node $v_i$ equipped with $M_i$ radio interfaces that can transmit on different channel simultaneously. We assume that at any given time a radio can only tune to one channel. Therefore, for successfully transmission a link $l$, both of nodes $b(l)$ and $e(l)$ must be tuned to one distinct channel $c$.

Let there are $F$ flows in the network. Each flow is associated with a source node and a destination node. Traffic for flow $f$ may be routed over multiple paths. Let $\mathbf{R}=[R]_{F\times L}$ represent routing matrix of network with 0 and 1 element and $R_{fl}=1$ denote flow $f$ traverse link $l$. Let $\vec{\lambda}=[\lambda_1,\lambda_2,...,\lambda_F]$ denote the offered load of the flows to the network and $\lambda_i$ is the traffic load that is generated by $i^{th}$ flow.

We assume that time is divided into slots of unit length, and the time that it takes to switch between channels is inconsiderable compared to the length of time slot. Each link l associated with a queue that denotes the number of packets queued at link l at the beginning of time slot t that we show by $q_l(t)$. The evolution of $q_l(t)$ may be written as:

$$Q_l(t+1) = Q_l(t) + \sum_{f\in F} R_{fl}.\lambda_f - r_l \quad (9)$$

### IV. ADAPTIVE CHANNEL ASSIGNMENT

Channel assignment in multi-radio wireless mesh networks involves allocating a channel for each mesh radio interface to achieve better channel utilization and decrease interference between channels. In other word we want to maximize throughput of system with better usage of resources such as channel and radio interfaces.

We propose a distributed learning automata based algorithms that dynamically adapt nodes channel selection according to load of the links. Each node equipped with a learning automaton that takes an action on each time slot τ. The action set of each automaton is set of subset of available channels that constructed by combination of $M_i$ (number of available radio for node i) from number of available channels (K).

First we define channel assignment matrix C as $CA=[c_{lk}]_{|L|\times|C|}$ such that $c_{lk}=1$ if channel $k$ assigned to link $l$. Our proposed algorithm estimates the available bandwidth that related to imposed interferences by neighbor nodes. If we divide $L(v_i)$ into two disjoint sets, $L_R(v_i)$ for receiving links and $L_S(v_i)$ for sending links, at the end of every timeslot, each node counts the number of packets received from 1-hop neighbor nodes. Let call it $A_l^i(t)$ and also $S_l^i(t)$ for number of packets sent on link $l$.

$$A_l^i = \begin{cases} \sum_{f \in F} R_{fl} . \lambda_f & \text{if } v_i = e(l) \\ 0 & \text{o.w} \end{cases} \quad (10)$$

$$S_l^i = \begin{cases} \sum_{f \in F} R_{fl} . \lambda_f & \text{if } v_i = b(l) \\ 0 & \text{o.w} \end{cases} \quad (11)$$

We use a simple channel state function for determine appropriateness of assigned channels in a node that reflect potential ability for sending available packets that received from 1-hop nodes.

$$CQ_i = \sum_{l' v_i = e(l')} \sum_{l v_i = b(l)} \frac{S_l^i(t)}{\max(A_l^i(t) + L(Q_l'), \theta)} R_{fl} . R_{fl'} \quad (12)$$

Multiplication of $R_{fl} . R_{fl'}$ ensures that $l$ and $l'$ are two consecutives link for flow f. $L(Q_l)$ is length of queued packets that could not be sent in the previous time slot. The fraction term imply quality of link l that when it is close to one, node can use bandwidth up to total nominal capacity of the link.

The minimum constant $\theta$ is a constant used to avoid division-by-zero anomalies when doesn't exist any packet for send. It is chosen to be much smaller than the typical values (currently $\theta = 1$). Each node responsible to maximize $S_l^i(t)$, so that channel state function ($CQ_i$) close to one. During the time slot, node records some statistical information about sent and received packets and the end of timeslot, automaton of node evaluate channel quality function and according to improvement or declination, automaton of node reward or penalized for current channel selection and respectively update channel set probability vector.

In this scheme each node tries to maximize its appropriateness function by maximizing of number of sent packets via current assigned channels and this will cause number of receiving packets increase for next 1-hop node. Nodes adaptively learn which channel-set is the best to improve quality of communication between neighbors. So in distributed fashion the overall throughput of the system will increase by appropriate channel assignment.

## V. STABILITY ANALYSIS

The aim of this section is to analyze the stability of learning method in the game of multi-automata. A play $a(t)=(a_1(t)...a_N(t))$ of n automata is a set of strategies chosen by the automata at stage $t$, such that $a^i(t)$ is the action selected by $i$th automata. Also $\beta(t)=(\beta_1(t)...\beta_N(t))$ is the payoff vector such that $\beta^i(t)$ is the payoff to $i$th automata.

Let the action set of all automata denoted by $A$ with $|A| = r$ such that $r$ is the number of combinations that we can assign the channels to available radios in a node.

Define function $F^i : \{(a_1,...,a_N)\} \to [0,1], \quad 1 \leq i \leq N$

$$F^i(a) = E\left[\beta_i \mid i^{th} \text{ automata chose } a_i \in A\right] \quad (13)$$

$F^i$ is the payoff function for player i. The players only receive the payoff such that they have no knowledge of these functions. We say $a^*=(a_1^*,...,a_N^*)$ is an optimal point of the game if for each i, $1 \leq i \leq N$,

$$F^i(a^*) \geq F^i(a),$$

for all $a = (a_1^*,...,a_{i-1}^*, a_i, a_{i+1}^*,...,a_N^*), a_i \neq a_i^*$

*REMARK 1* In the above definition, the condition implies that $a^*$ is a Nash equilibrium of the game matrix $F(.)$ indexed by $a_i$, $1 \leq i \leq N$.

Two known theorems that we used for the stability of systems are stated as follow:

*Theorem 1:* Consider the discrete-time system

$$X(t+1)=f(X(t)) \quad (14)$$

Where $X$ is a vector, $f$ is a vector such that $f(0)=0$. Suppose there exist scalar function $V(x)$ continuous in $x$ such that:

(1) $V(X) > 0$ for $X \neq 0$.
(2) $\Delta V(X) < 0$ for $X \neq 0$.
(3) $V(0)=0$.
(4) $V(X) \to \infty$ as $\|X\| \to \infty$.

Then equilibrium state $X=0$ is asymptotically stable and $V(X)$ is a Lyapunov function.

*Theorem 2:* if the function $f(X)$ defined as:

$$\|f(X)\|_1 < \|X\|_1 \text{ with } f(0) = 0 \quad (15)$$

for some set of $X \neq 0$ the system above is asymptotically stable and one of its Lyapunov function is :

$$V(X) = \|X\|_1 \qquad (16)$$

These two theorems and their proofs are given in [18].

*Theorem 3:* A multi-player game of automata using the proposed learning scheme in stationary environment reach the pure optimal strategy.

An Identical analysis for single automaton has been proposed in [19]. In this study easily shown that the 1$^{st}$ norm of expected value of all sub-optimal action converges to zero. For example for $i$th automaton if $a_1$ is optimal action and $\overline{P_i}(t+1)$ is a vector of expected value for other sub optimal action in steady state, we can show that:

$$\|\overline{P}(t+1)\|_1 < \|\overline{P}(t)\|_1,$$

We know that every automaton in the network choose an action independently base on its action probability vector, so we can extend this analysis to matrix of all probability vectors. So if we define $f(\overline{P}(t)) = \overline{P}(t+1)$ and extending $P(t)$ to matrix of all probability vectors of every automata, we have proven that, the expected values of the probabilities of the sub-optimal actions all converge to zero. This implies that the probability of the optimal action converges to 1, and according to Lyapunov function the stability of the optimal point is proven [17].

## VI. EXPERIMENTAL RESULTS

In this section, we evaluate the performance of our learning automata based channel assignment scheme through detailed numerical studies. We modified JiST/SWANS [15] to support multi-radio multi-channel wireless mesh networks, so each node can use multi radio network interface card and can transmit packets concurrently on its radio interface that assigned by different channels.

A grid wireless mesh network topology generated using the following method. We first specify a square region with the area of 1500×1500 that has the width [0, 1500] on the x axis and the height [0, 1500] on the y axis. Then we generate a certain number of nodes and placed in the same distance from each other. We assume that each end user has infinite number of packets for send. We use Rayleigh for Fading model and free space for path loss. All of mesh routers are fixed and only client nodes can be mobile. Default radio reception sensitivity and radio reception threshold are subsequently assumed -91dBm and -81dBm and also for simplicity ambient noise was ignored. The total simulation time for each scenario is 200 seconds. First 50 seconds that allow simulations to reach a stable state, is not used for evaluation.

The network topologies used in our experiments is grid style is depicted in Fig 1. Each node equipped with a learning automaton that evaluates network dynamics according to total delivered packets. The learning automaton of each node randomly select R channel from K=10 orthogonal channels in the first time slot. After channel allocation, nodes start sending and receiving packets like a heartbeat application. At the end of the each time slot total number of packets that delivered by each node registered as network response to channel assignment. At the beginning of each time slot, nodes evaluate environment response to last channel-set allocation and reward/penalize their automata by updating action probabilities.

In the first scenario we equipped each node subsequently with single, two and three 802.11a radio interface cards with distance-rate relationships commonly advertised by 802.11a vendors. The potential rate of the link is 1Mbps and default radio frequency is set to 2.4GHz. Radio reception sensitivity, radio reception threshold and transmission strength are set to -91dB, -81dB and 15 dB respectively. For simplicity, the ambient noise and antenna gain was not considered. Fig. 2 illustrates the performance of our algorithm by total delivered packets. Using more radio interface card in each node, results significant improvement in total delivered packets as 1.7 times according to double radio and 4 times according to single radio. Total network dynamics in long time period show that system can efficiently adapt their behavior according to environment response, despite to minor oscillations in delivered packets in short time periods, network can efficiently learn how choose appropriate channels that deliver more packets with low interference and more connectivity.

Fig. 3 and 4 shows the impact of our algorithm on network connectivity and total interference imposed by channel allocation scheme.

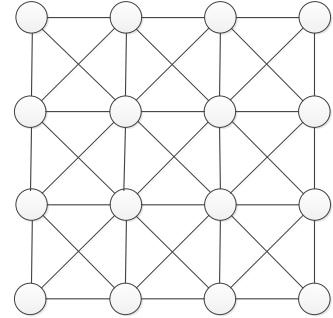

Figure 1. Grid style topology used in simulation

Network connectivity as the number of available common channel between nodes on transmission region directly depends on how we assign channels to links. According to Fig. (3), Network connectivity is very poor on beginning of our algorithm. This is because of blind randomly assignment of channels to nodes in the early time slots, so after the few iteration connectivity of the network dynamically improve fast. In three radio mode, chance of assigning of common channel between nodes is high, so in the early time slots, system improve the connectivity very fast.

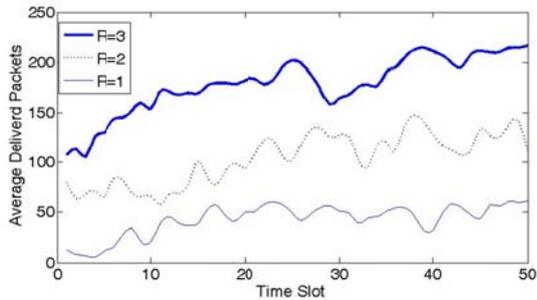

Figure 2. Average delivered packets by nodes in 50 time slot

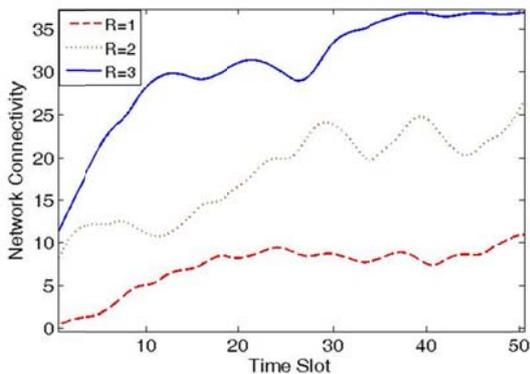

Figure 3. Network connectivity improvement

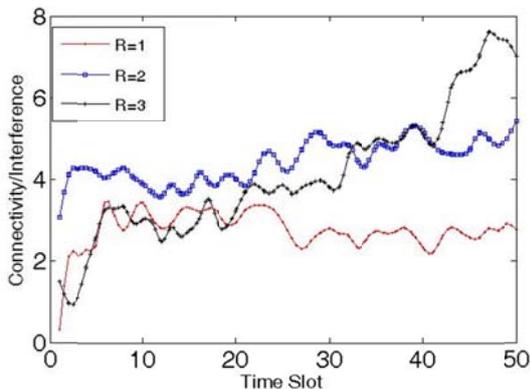

Figure 4. Effect of algorithm on connectivity and Interference

Investigating the interference property can offer good insight into the operation of our algorithm. Fig. 4 clearly indicates evolution of interference according to network connectivity across the time slots. Obviously by increasing of network connectivity, potential interference will be increased, so strongly connected network has high potential interference. To measure the quality of the solution in term of channel interference, we track the behavior of the system by connectivity growing beside of interference increasing. Increasing in proportional connectivity vs. interference shows that algorithm effectively decreases interference and increases network topology connectivity proportionally.

VII. CONCLUSIONS

In this paper, we proposed learning automata based distributed and online channel assignment for multi–radio, multi-channel wireless mesh networks. According to unknown characteristics of wireless mesh networks, mesh routers that equipped by learning automata can predict link quality with environment response of the networks for last behavior of network components. This distributed fashion of online channel allocation leads the network to achieve optimum point of overall network throughput. We analyze this mechanism by expecting of sub-optimal action probabilities, that stability theory was used to characterize the dynamic behavior of our approach. In addition with Lyapunov-like function, the stability and optimum point converge under some mild restriction. The experimental results show that our scheme provides significant improvement in network behavior and it can establish good balance for network connectivity and interference.